\documentclass[a4paper]{jpconf}
\usepackage{graphicx}
\usepackage{amsfonts}
\usepackage{subfigure, wrapfig}
\begin{document}

\title{M\small{AJORANA} Collaboration's Experience with Germanium Detectors}
\author{S.~Mertens$^{1}$,
N.~Abgrall$^{1}$, 
F.~T.~Avignone III$^{2,3}$, 
A.~S.~Barabash$^{4}$, 
F.~E.~Bertrand$^{3}$, 
V.~Brudanin$^{5}$, 
M.~Busch$^{6,7}$, 
M.~Buuck$^{8}$,
D.~Byram$^{9}$,
A.S.~Caldwell$^{10}$,
Y-D.~Chan$^{1}$, 
C.~D.~Christofferson$^{10}$, 
C.~Cuesta$^{8}$, 
J.~A.~Detwiler$^{8}$, 
Yu.~Efremenko$^{11}$, 
H.~Ejiri$^{12}$, 
S.~R.~Elliott$^{13}$, 
A.~Galindo-Uribarri$^{3}$, 
G.~K.~Giovanetti$^{14,7}$, 
J.~Goett$^{13}$,
M.~P.~Green$^{3}$, 
J.~Gruszko$^{8}$,
I.~Guinn$^{8}$,
V.~E.~Guiseppe$^{2}$, 
R.~Henning$^{14,7}$, 
E.~W.~Hoppe$^{15}$, 
S.~Howard$^{10}$, 
M.~A.~Howe$^{14,7}$, 
B.~R.~Jasinski$^{9}$, 
K.~J.~Keeter$^{16}$, 
M.~F.~Kidd$^{17}$, 
S.~I.~Konovalov$^{4}$, 
R.~T.~Kouzes$^{15}$, 
B.~D.~LaFerriere$^{15}$, 
J.~Leon$^{8}$, 
J.~MacMullin$^{14,7}$, 
R.~D.~Martin$^{9}$, 
S.~J.~Meijer$^{14,7}$,
J.~L.~Orrell$^{15}$, 
C.~O'Shaughnessy$^{14,7}$,
N.~R.~Overman$^{15}$, 
A.~W.~P.~Poon$^{1}$, 
D.~C.~Radford$^{3}$, 
J.~Rager$^{14,7}$,
K.~Rielage$^{13}$, 
R.~G.~H.~Robertson$^{8}$, 
E.~Romero-Romero$^{11,3}$,
M.~C.~Ronquest$^{13}$, 
B.~Shanks$^{14,7}$,
M.~Shirchenko$^{5}$, 
N.~Snyder$^{9}$,
A.~M.~Suriano$^{10}$,
D.~Tedeschi$^{2}$,
J.~E.~Trimble$^{14,7}$,
R.~L.~Varner$^{3}$, 
S.~Vasilyev$^{5}$,
K.~Vetter$^{1}$ \footnote[18]{Alternate Address: Department of Nuclear Engineering, University of California,
Berkeley, CA, USA},
K.~Vorren$^{14,7}$, 
B.~R.~White$^{3}$,
J.~F.~Wilkerson$^{14,7,3}$, 
C.~Wiseman$^{2}$,
W.~Xu$^{13}$,
E.~Yakushev$^{5}$, 
C-H.~Yu$^{3}$,
and V.~Yumatov$^{4}$\\The M\small{AJORANA} Collaboration}

\address{$^{1}$Nuclear Science Division, Lawrence Berkeley National Laboratory, Berkeley, CA, USA}
\address{$^{2}$Department of Physics and Astronomy, University of South Carolina, Columbia, SC, USA}
\address{$^{3}$Oak Ridge National Laboratory, Oak Ridge, TN, USA}
\address{$^{4}$Institute for Theoretical and Experimental Physics, Moscow, Russia}
\address{$^{5}$Joint Institute for Nuclear Research, Dubna, Russia}
\address{$^{6}$Department of Physics, Duke University, Durham, NC, USA}
\address{$^{7}$Triangle Universities Nuclear Laboratory, Durham, NC, USA}
\address{$^{8}$Center for Experimental Nuclear Physics and Astrophysics and \\
             Department of Physics, University of Washington, Seattle, WA, USA}
\address{$^{9}$Department of Physics, University of South Dakota, Vermillion, SD, USA}
\address{$^{10}$South Dakota School of Mines and Technology, Rapid City, SD, USA}
\address{$^{11}$Department of Physics and Astronomy, University of Tennessee, Knoxville, TN, USA}
\address{$^{12}$Research Center for Nuclear Physics and Department of Physics, Osaka University, Ibaraki, Osaka, Japan}
\address{$^{13}$Los Alamos National Laboratory, Los Alamos, NM, USA}
\address{$^{14}$Department of Physics and Astronomy, University of North Carolina, Chapel Hill, NC, USA}
\address{$^{15}$Pacific Northwest National Laboratory, Richland, WA, USA}
\address{$^{16}$Department of Physics, Black Hills State University, Spearfish, SD, USA}
\address{$^{17}$Tennessee Tech University, Cookeville, TN, USA}

\ead{smertens@lbl.gov}

\begin{abstract}
The goal of the \textsc{Majorana} \textsc{Demonstrator} project is to search for 0$\nu\beta\beta$ decay in $^{76}\mathrm{Ge}$. Of all candidate isotopes for 0$\nu\beta\beta$, $^{76}\mathrm{Ge}$ has some of the most favorable characteristics. Germanium detectors are a well established technology, and in searches for 0$\nu\beta\beta$, the high purity germanium crystal acts simultaneously as source and detector. Furthermore, p-type germanium detectors provide excellent energy resolution and a specially designed point contact geometry allows for sensitive pulse shape discrimination. This paper will summarize the experiences the \textsc{Majorana} collaboration made with enriched germanium detectors manufactured by ORTEC$^{\circledR}$. The process from production, to characterization and integration in \textsc{Majorana} mounting structure will be described. A summary of the performance of all enriched germanium detectors will be given. 
\end{abstract}

\section{Introduction}
A unique way to explore the nature of the neutrino is the search for neutrinoless double beta decay (0$\nu\beta\beta$)~\cite{Goe35, Maj37, Avi08}. Observation of 0$\nu\beta\beta$-decay would decisively prove that neutrinos are Majorana particles and that lepton number is violated. The \textsc{Majorana} \textsc{Demonstrator} will perform a search for 0$\nu\beta\beta$-decay in $^{76}\mathrm{Ge}$~\cite{MJD14,Wenqin}. The experiment is currently under construction at the Sanford Underground Facility (SURF) in South Dakota, USA. 

The \textsc{Majorana} \textsc{Demonstrator} will use an array of 40~kg of high purity germanium detectors, up to 30~kg of which will be enriched to 87\% in $^{76}\mathrm{Ge}$, surrounded by passive and active shielding. The major goal is to demonstrate a path forward to achieving a background rate at or below 3 counts/(ROI-t-y) in the 4~keV region of interest (ROI) around the 2039~keV Q-value of the $^{76}\mathrm{Ge}$ 0$\nu\beta\beta$ decay. This is required for the next generation of tonne-scale germanium-based 0$\nu\beta\beta$-decay searches that are designed to probe the neutrino mass scale in the inverted-hierarchy region.

Beyond the usage of ultra-clean material and extensive shielding, a key feature to achieve a low background level is a high energy resolution to minimize the ROI and pulse-shape discrimination capability to distinguish signal from background events. These properties are provided by High Purity (HP) P-type Point Contact (PPC) germanium (Ge) detectors. By enriching the natural germanium to 87\% in $^{76}\mathrm{Ge}$ the detectors can be used as source and detector simultaneously. At the time of this presentation, 30 enriched detectors with average mass of 840~g have been delivered and successfully tested at SURF. All detectors met the requirements during characterization in vendor cryostat and are now being assembled in the final \textsc{Majorana} \textsc{Demonstrator} experiment.

\section{Production}
The \textsc{Majorana} enriched $^{76}\mathrm{Ge}$ material was produced by the Electrochemical Plant (ECP) in Zelenogorsk, Russia. ECP has a long history of supplying large quantities of enriched $^{76}\mathrm{Ge}$ material to the world scientific community, e.g.\ to the GERmanium Detector Array (GERDA) Experiment~\cite{Gerda}. A total of about 42.5~kg of enriched $^{76}\mathrm{Ge}$ material has been purchased and provided for the \textsc{Majorana} \textsc{Demonstrator}. The germanium was isotopically enriched in $^{76}\mathrm{Ge}$ from 7.8\% to 87\% through centrifuging germanium tetrafluoride ($^{76}\mathrm{GeF_4}$) and then converted to germanium oxide ($^{76}\mathrm{GeO_2}$).

\begin{figure}
  \centering
  \subfigure[]{\includegraphics[width = 0.45\textwidth]{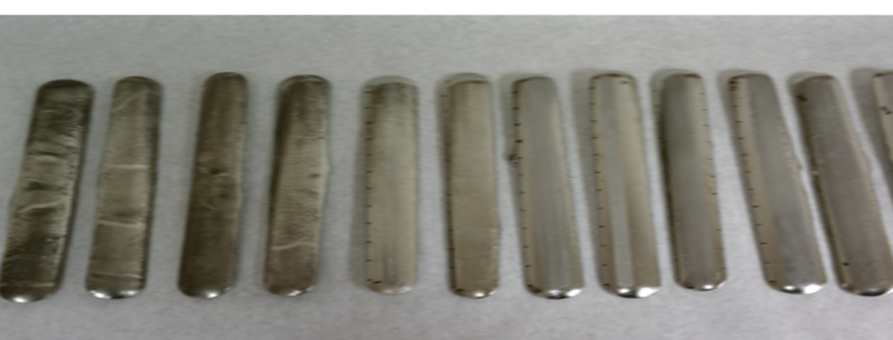}}
  \subfigure[]{\includegraphics[width = 0.45\textwidth]{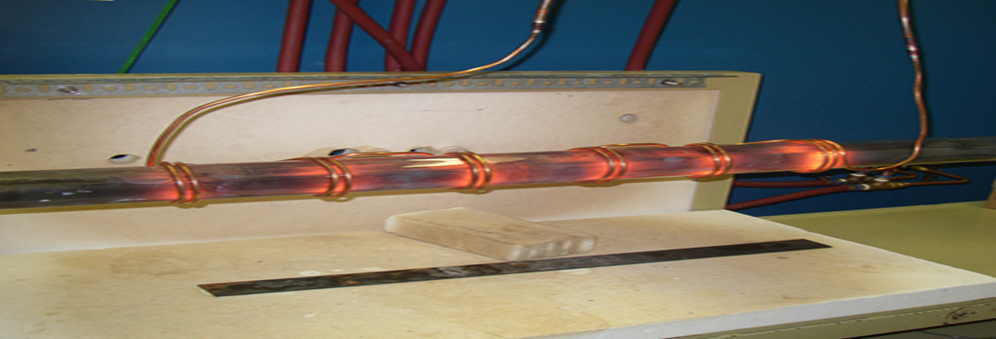}}
  \caption{(a): Ge metal bars after reduction to $^{76}\mathrm{Ge}$ from $^{76}\mathrm{GeO_2}$. (b): Zone refinement setup at ESI.}
 \label{fig:Production}
\end{figure}

All material in the form of $^{76}\mathrm{GeO_2}$ was then shipped to Oak Ridge, TN, USA. Transportation in a special steel-shielding container reduced the activation of $^{68}\mathrm{Ge}$ and $^{60}\mathrm{Co}$ due to cosmic-ray exposure by factors of 10 and 15, respectively. The total effective (i.e.\ equivalent exposure at sea level) exposure including the shipment and enrichment is estimated to be 12.4~days. At Electrochemical Systems Inc. (ESI) in Oak Ridge, Tennessee the $^{76}\mathrm{GeO_2}$ material was reduced via high temperature reduction in a hydrogen atmosphere. Subsequently, the material was zone-refined to electronic grade (${\sim}10^{13}$ net impurities per cubic centimeter) bars, see Fig.~\ref{fig:Production}. The yield at ESI for all 42.5~kg of $^{76}\mathrm{Ge}$ was $>98$\%. At the time of this presentation, scrap material is being recycled, to further increase the yield.

At ORTEC$^{\circledR}$~\cite{ortec} the manufacturing of the detectors took place after a second zone refinement to ${\sim}10^{11}$ net impurities per cubic centimeter. Before the shipment to SURF the first characterization campaign was performed at ORTEC$^{\circledR}$. By storing the germanium in a cave with an overburden of 40~m of rock, located about 8~km from the processing facility, a total effective exposure at ESI and ORTEC$^{\circledR}$ of ${\sim}14$ days was estimated. Finally, the detectors were transported to SURF by car, leading to an additional exposure time of ${\sim}35$~h. Overall, the total effective sea-level equivalent exposure is approximately a factor of a few better than the project specification of no more than 100 days. At the time of this presentation, 25.2~kg of enriched germanium detectors are underground and \textsc{Majorana} is aiming for additional ${\sim}4$~kg of detectors from the recovery of scrap material.

\section{Characterization in Vendor Cryostats}
All enriched detectors are extensively tested and characterized in their vendor cryostat both at ORTEC$^{\circledR}$ and at SURF. The tests include measurements of the mass, impurity concentration, depletion and operating voltages, leakage current, energy resolution, electronic noise, dead layer, relative efficiency compared to a 3x3-inch NaI detector, and pulse-shape discrimination performance. In this section the basic working principle of PPC Ge detectors with respect to these properties is outlined, the main characterization results are presented, and details of pulse-shape discrimination performance are discussed.

\subsection{PPC geometry}
\begin{figure}
  \centering
  \begin{minipage}{0.6\textwidth}
   \includegraphics[width = \textwidth]{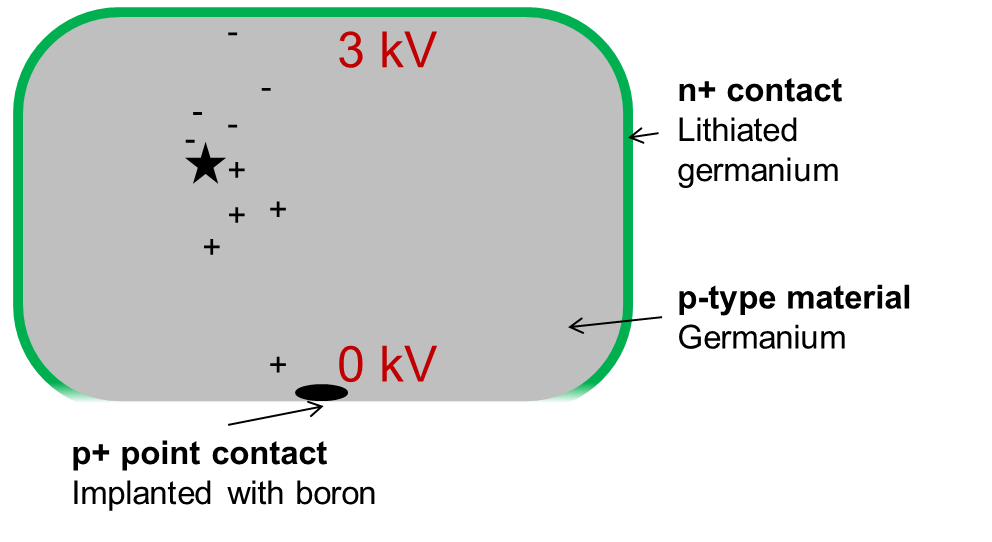}
  \end{minipage}
  \hfill
  \begin{minipage}{0.39\textwidth}
    \caption{Schematic view of a PPC Ge detector. The green line indicates the lithiated $\mathrm{n}^+$-type contact, which is held at positive bias voltage. Here 3~kV is chosen as a typical example. The small point contact is illustrated as a black oval point at the bottom of the detector, it is kept at zero potential. In this configuration the holes created by an energy deposition in the bulk of the detector move to the point contact, whereas the electrons move to the $\mathrm{n}^+$-type contact.}
    \label{fig:PPC}
  \end{minipage}
\end{figure}

The bulk of the detector is p-type germanium. P-type material is advantageous for efficient charge collection and therefore high energy resolution. Furthermore, it is preferable with respect to ease of handling and blocking of surface alphas. The lithiated surface of the detector provides the $\mathrm{n}^+$-type contact, and a small $\mathrm{p}^+$ point contact is created via boron implantation. The $\mathrm{n}^+$-type contact is held at positive bias voltage, whereas the point contact is kept a zero potential, see Fig.~\ref{fig:PPC}. The surface between the point contact and the $\mathrm{n}^+$-type surface is covered with a passivation layer and is one of the most sensitive parts of the detector, any damage or contamination on this surface can lead to large leakage currents and malfunctioning of the detector. 

The particular geometry of a PPC detector creates a weighting potential, such that the leading edge of the signal is largely created only when the holes arrive at the immediate vicinity of the point contact. This leads to pulse shapes which are almost independent of their point of origin, and furthermore creates distinct deviations from this shape when an event is composed of energy depositions at multiple positions in the detector. More details of the pulse shape discrimination capability of PPC detector are discussed in section~\ref{sec:pulse}.

\subsection{General performance results}
In order to maximize the exposure and keep the number of channels at a manageable level, large mass detectors are preferable. At the same time, in order to fully deplete the detectors at a technically feasible operating voltage, the impurity concentration must be kept at a low level. With an average mass of 840~g and an average impurity concentration of $7\cdot10^{9}$ net impurities per cubic centimeter all \textsc{Majorana} enriched detectors are operated at a bias voltage of less than 5~kV. Furthermore, the average leakage current of all detectors is only ${\sim}$11~pA at full bias voltage.

\begin{figure}
  \centering
  \subfigure[]{\includegraphics[width = 0.49\textwidth]{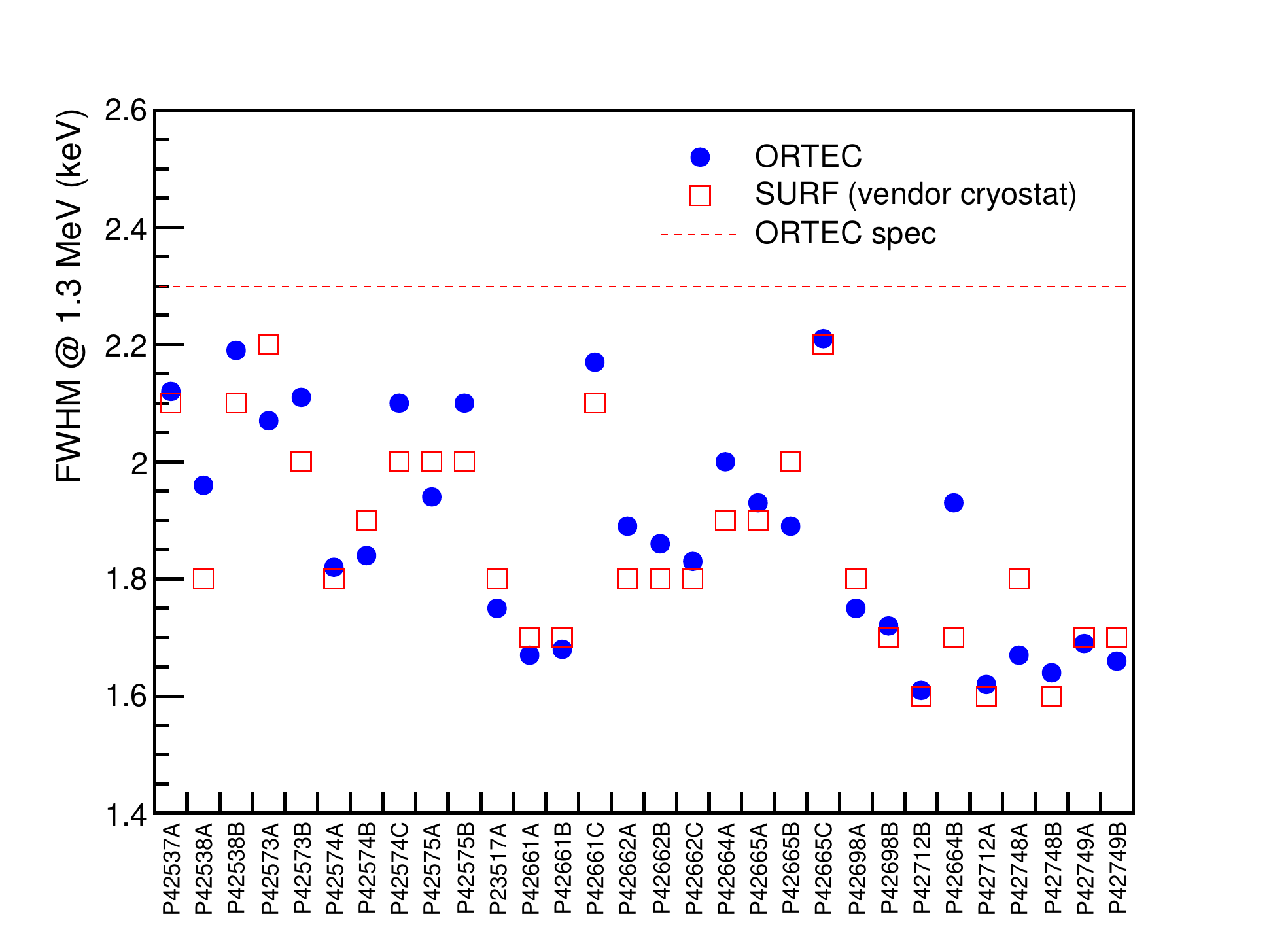}}
  \subfigure[]{\includegraphics[width = 0.49\textwidth]{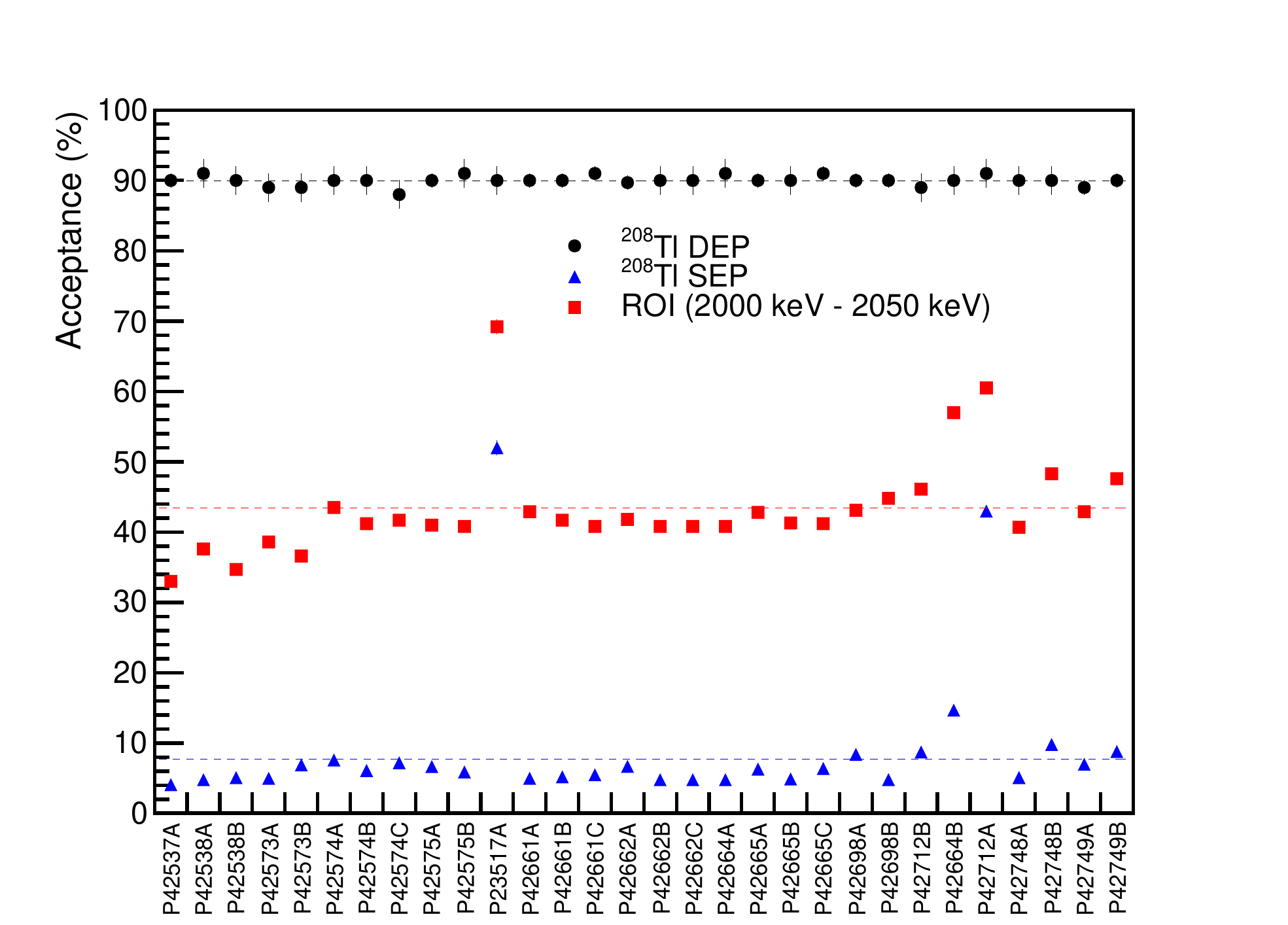}}
  \caption{(a): Energy resolution of all 30 enrichted detectors, measured both at SURF and ORTEC$^{\circledR}$. All detectors meet the specification. (b): Pulse shape discrimination performance of all 30 enriched detectors. By fixing the acceptance in the double-escape peak (DEP) to 90\%, the counts in the single-escape peak can be reduced to less than 10\% for all detectors but three. All detector will still be used in the final \textsc{Majorana} \textsc{Demonstrator} experiment. The degraded performance of P23517A can likely be related to its very low impurity concentration, see Section~\ref{ssc:PSAimpurity}. A flat edge of the crystal might explain the pulse-shape discrimination behaviour of P42664B and the performance of P42712A can be explained by an instable depletion voltage during the measurement.}
 \label{fig:characterization}
\end{figure}

A major advantage of P-type point contact detectors is their excellent energy resolution, which minimizes the energy width of the ROI, and maximizes the signal-to-background ratio. The resolution is typically measured via the FWHM of the 1.3~MeV line of a $^{60}\mathrm{Co}$ calibration source. With an average FWHM of 1.9~keV at 1.3~MeV, all 30 detectors exceed the vendor specification of 2.3~keV, see Fig.~\ref{fig:characterization}. ORTEC$^{\circledR}$ and SURF measurements are in very good agreement. At the 122~keV line of $^{57}\mathrm{Co}$ an average FWHM of 624~eV is achieved at ORTEC$^{\circledR}$.

A low electronic noise level allows for a low energy threshold. This is preferable for example to assure a reliable rejection of background due to cosmogenically induced $^{68}\mathrm{Ge}$. 
This background is expected to be reduced to 0.14 counts/ROI/t/y averaged over the first year, by applying an single-site time correlation cut for five $^{68}\mathrm{Ga}$-half-lives following 10-keV K-shell deexcitations. If the ~1-keV L-shell de-excitations can also be tagged, this background may be reduced by another order-of-magnitude. 

Furthermore, a low energy threshold allows for the option of extending the physics reach of \textsc{Majorana} to a search for new physics, such as low-mass WIMPS~\cite{Wimp}, which would leave a signal at the low energy end of the spectrum. The electronic noise is determined at ORTEC$^{\circledR}$ via the FWHM of a pulser signal. All detectors meet the requirement of $\mathrm{FWHM}_{\mathrm{pulser}}<500$~eV.

To maximize the active volume of the detectors a thin dead layer is required. On the other hand, a minimal thickness of at least 0.01~mm~\cite{alpha} is advantageous, as it prevents surface alphas to deposit energy in the ROI. With an average dead layer of 0.8~mm all 30 enriched detectors meet the requirements. The dead layer has been determined by comparing the ratio of the peak heights of 81~keV and 356~keV $^{133}\mathrm{Ba}$ lines to simulation to determine the relative difference in attenuation due to the inactive Ge. Measurements at ORTEC$^{\circledR}$ and SURF are in good agreement.

The relative efficiency compared to a 3 $\times$ 3 inch NaI(Tl) detector is determined with a $^{60}\mathrm{Co}$ source of known strength, which is placed at a given distance to the detector, on axis with the center of the crystal. The counts in the photo peak at 1.33~MeV are measured and then normalized to the efficiency of a standard 3x3-inch NaI detector. The measured efficiencies agree with what ORTEC$^{\circledR}$ measured and our simulation. Measurements at ORTEC$^{\circledR}$ and SURF are in good agreement.

\subsection{Pulse shape discrimination performance}
\label{sec:pulse}
\begin{figure}
  \centering
  \subfigure[]{\includegraphics[width = 0.49\textwidth]{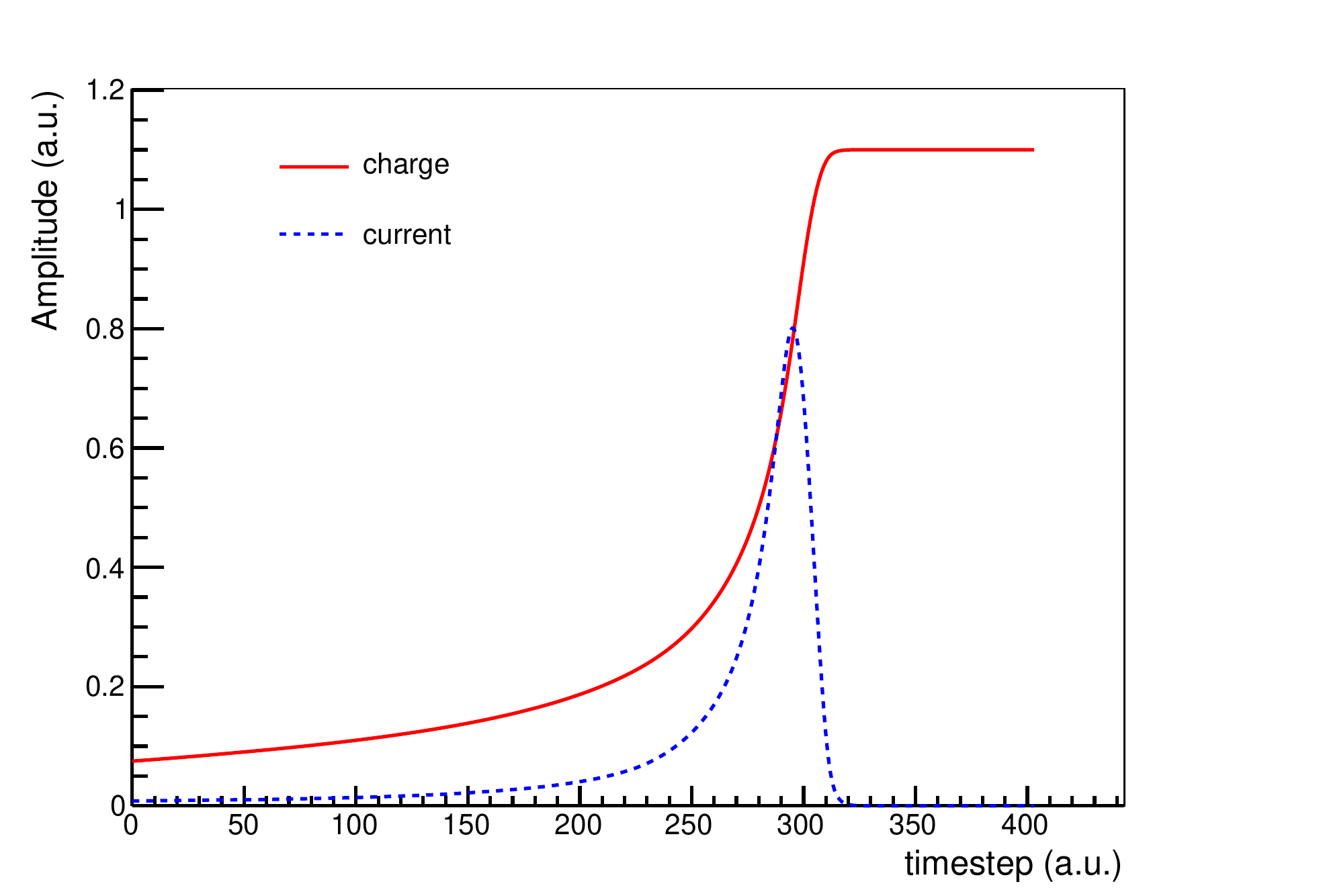}}
  \subfigure[]{\includegraphics[width = 0.49\textwidth]{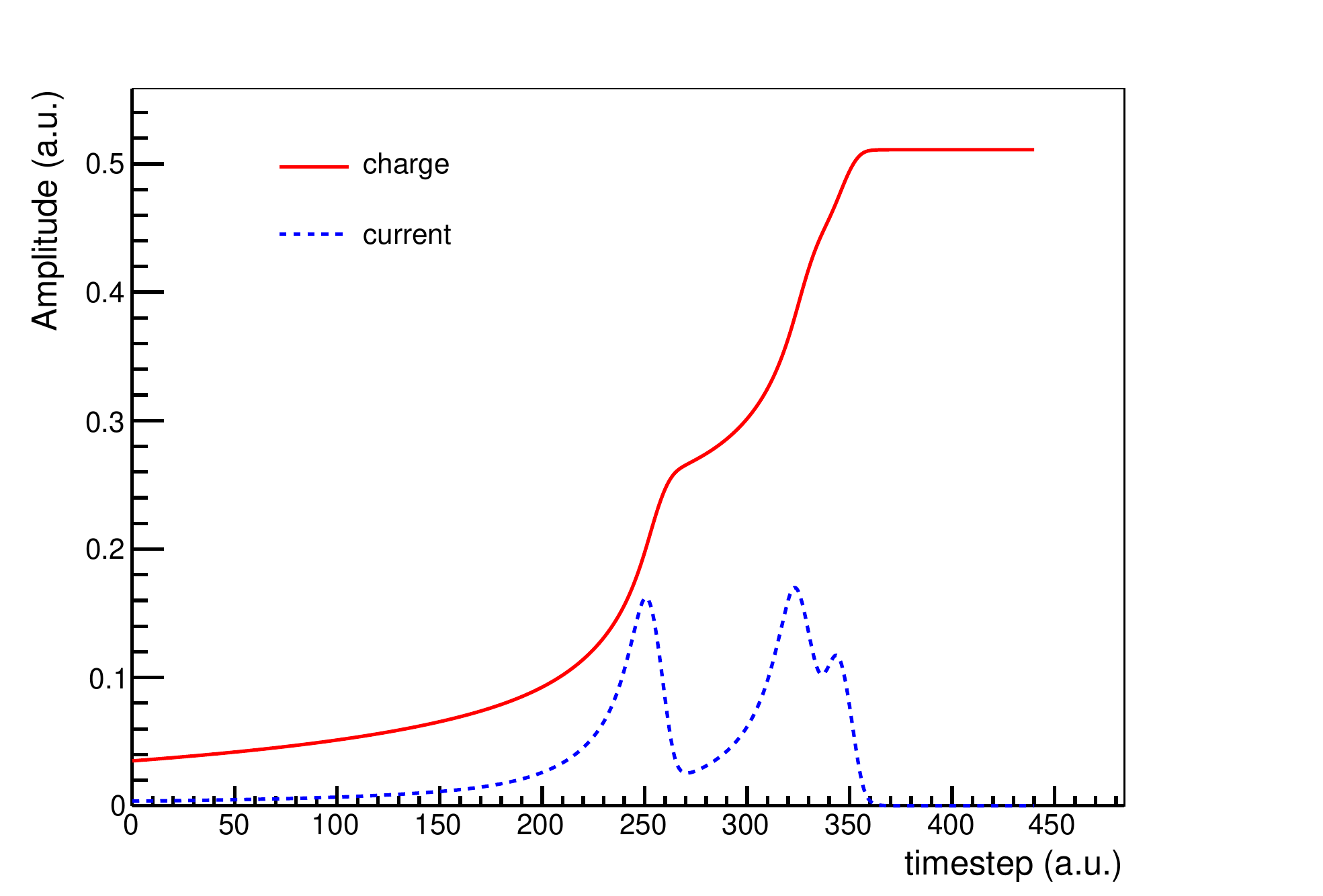}}
  \caption{(a): Simulation of a single-site event. (b): Simulation of multiple-site event. The red solid line shows the charge as as a function of time, the blue dotted line corresponds to its derivative, i.e.\ the current. The 10\% to 90\% rise time of the signal is created only while the charge is in very close vicinity to the point contact, see Fig.~\ref{fig:field}. Consequently, the ratio of maximum current to the amplitude of the signal (energy) is very similar for all single-site events and different for multiple-site events.}
 \label{fig:SSEMSE}
\end{figure}

A key feature of the point contact geometry is its capability of pulse-shape discrimination. The two electrons emitted in 0$\nu\beta\beta$ deposit their energy in a small volume of the detector and are therefore referred to as single-site event. A large class of background events is due to gammas which have high probability to deposit their energy at multiple positions in the detectors, and are therefore called multi-site events. The point-contact geometry creates a weighting potential in the detector such that the pulse shape of single-site events is significantly different from multi-site events, see Fig.~\ref{fig:SSEMSE}. A simple discriminative property of the pulse shape is the ratio of the maximum current (A) and the amplitude (E) of the pulse. 

To determine the effectiveness of single-site and multi-site event discrimination, a ${}^{228}\mathrm{Th}$ calibration source is used. In the decay of ${}^{208}\mathrm{Tl}$ a high energy (2.615~MeV) gamma is produced, which has a high probability of producing an $e^{+}$ $e^{-}$ pair in the Ge detector. The subsequent two gammas produced by annihilation of the positron can either fully deposit their energy in the detector (full energy peak), only one gamma deposits energy in the detector (single escape peak), or both gammas leave the detector without interaction (double escape peak). The latter generates a typical single site event, whereas the single escape peak is populated with multi-site events. These two event classes are ideal to evaluate the efficiency of the pulse shape discrimination. 

By fixing the acceptance of single-site events to 90\%, the cut efficiency for the single-escape peak and the ROI is determined. For all enriched detectors, with the exception of one, see Fig.~\ref{fig:characterization}, the single-escape events could be reduced to less than 10\%. The number of events in the ROI is reduced to about 40\%, which corresponds to the expected number of single-site events in this energy range. 

\subsection{Impact of ultra-low impurity concentrations on pulse-shape discrimination performance}
\label{ssc:PSAimpurity}
Detector P23517A which showed the poorest pulse-shape discrimination performance (see Fig.~\ref{fig:characterization}) also exhibits the lowest depletion voltage, which can be related to an extremely low impurity concentration. Dedicated simulations with the simulation software \textit{siggen}~\cite{siggen} reveal a clear impact of ultra-low impurity concentrations on the pulse-shape discrimination performance. 

\begin{figure}
  \centering
  \subfigure[]{\includegraphics[width = 0.49\textwidth]{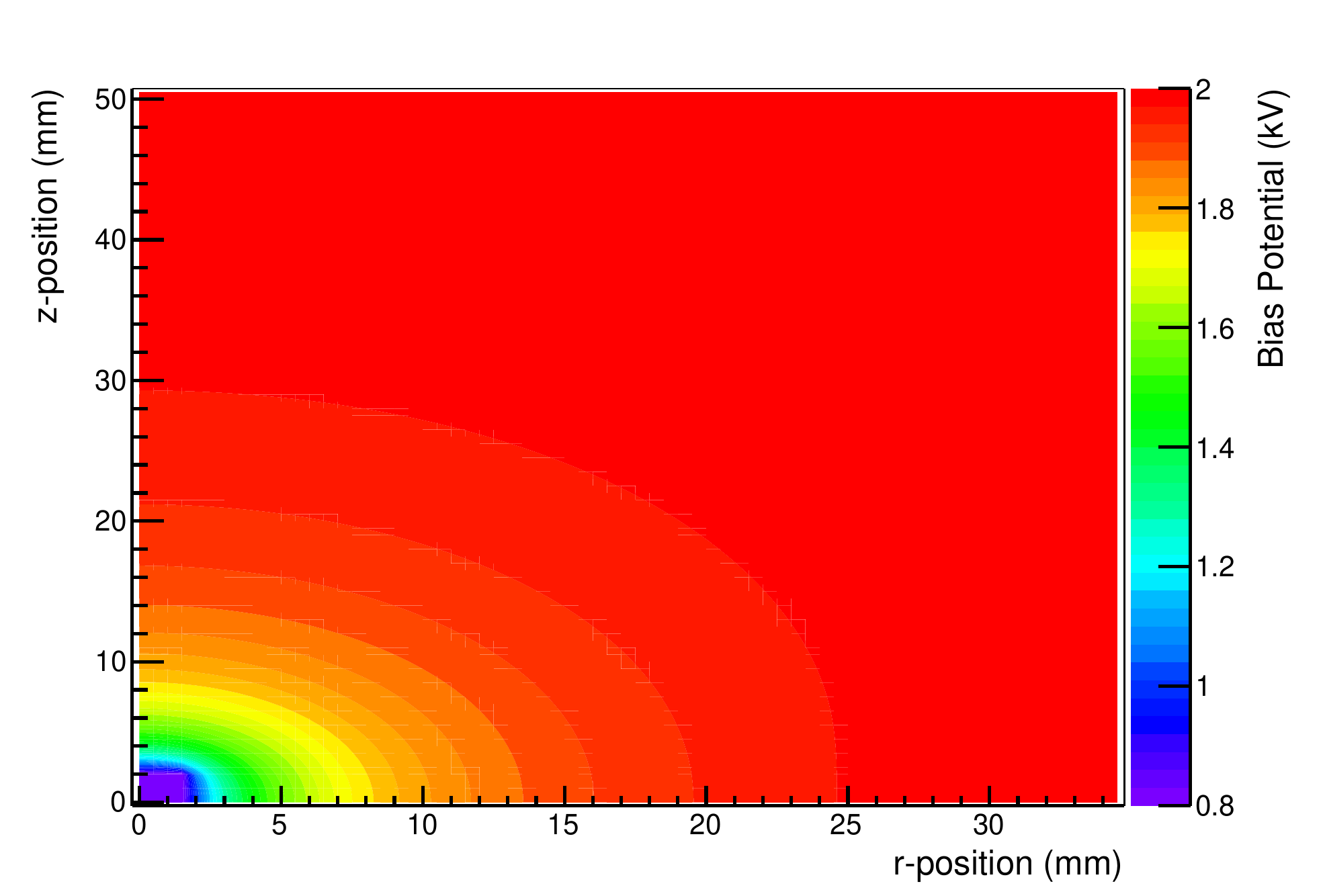}}
  \subfigure[]{\includegraphics[width = 0.49\textwidth]{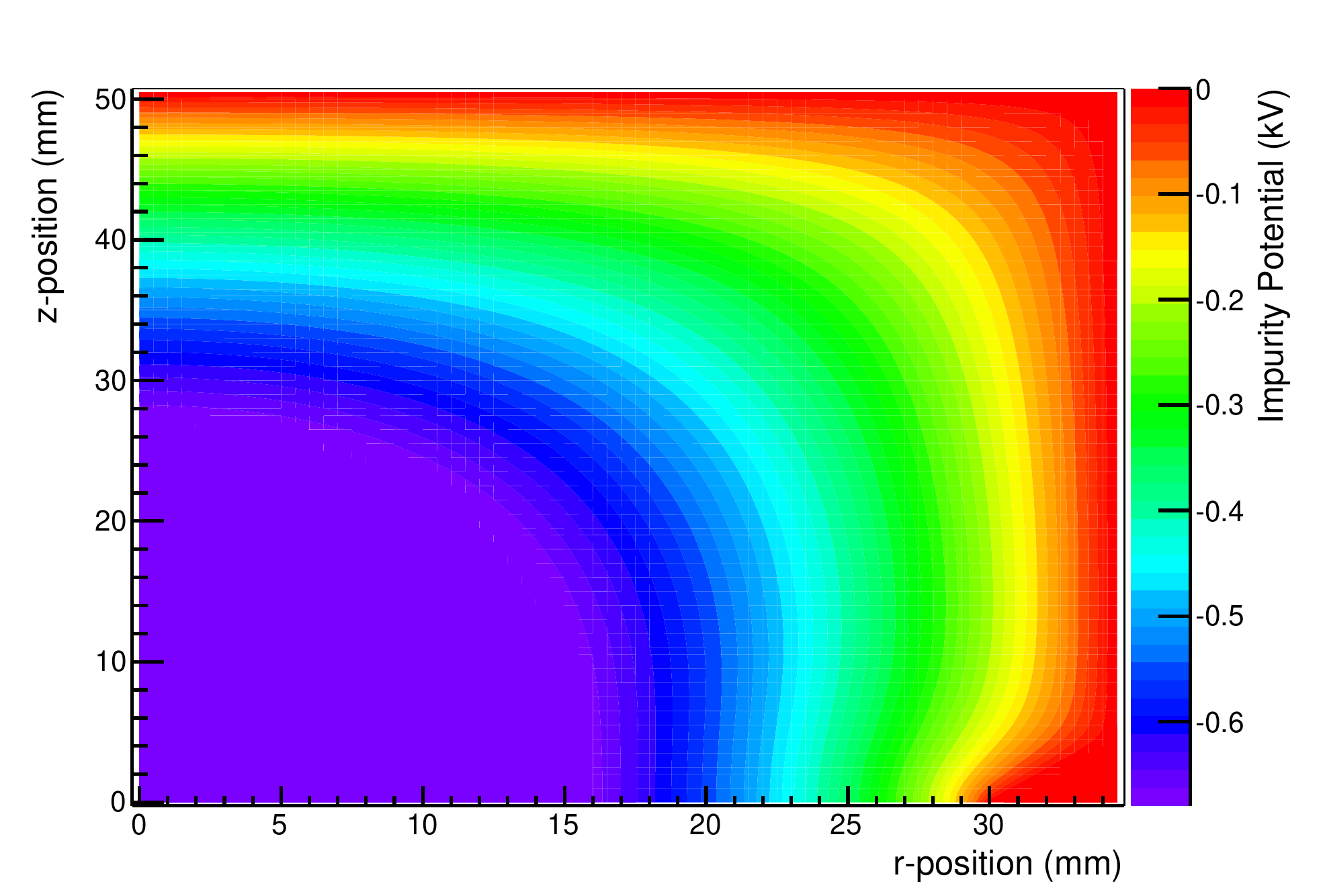}}
  \caption{(a): Electric potential created by the bias voltage. In this example the bias is voltage is 3~kV. This potential determines the weighting potential of the detector. (b): Electric potential created by the impurity concentration. At the conductive surfaces of the detector the potential is forced to zero volts, whereas the high resistive passivation layer can be at a finite potential. In this example model of a detector with an impurity concentration of $3.2\cdot10^9$ net impurities per cubic centimiter at the top of the detector and an impurity gradient of $0.25\cdot10^9$ change in impurity density per centimiter was assumed. The total electric potential in the detector is given by the sum of the potentials in a) and b).}
 \label{fig:field}
\end{figure}

This fact can be understood by considering the electric field in a detector. The total electric potential is given as a superposition of the potential created by the bias voltage and the potential created by the impurity concentration. In particular, the field in the corners is governed by the impurity concentration, as can be seen in Fig.~\ref{fig:field}. To first order, the time difference $\Delta t \approx \frac{\Delta x}{v_{\mathrm{init}}}$ between the charges arriving at the point contact is given by their initial spatial separation $\Delta x$ relative to their initial velocity $v_{\mathrm{init}}$. As the velocity is determined by the electric field, $\Delta t$ and correspondingly the maximum current A, is small for charges created in the corners of an ultra-low impurity.

The spatial separation is driven by diffusion and self-repulsion. Analytic estimates show that, while charges move in the low field region (in low-impurity detectors the field can drop below 50 V/cm in the corners), these effects can cause a spread of the charges of up to about 1~mm. As a first approximation, the spatial separation of charges is implemented in the simulation code by a convolution of the simulated signal by a Gaussian with a width corresponding to the charge cloud size.

To validate the model a set of dedicated tests with a natural ultra-low purity detector with approximately $0.5\cdot10^9$ net impurities per cubic centimeter has been performed. First, the radial dependence of the cut parameter A/E was determined with the help of a radial scan using an $^{241}\mathrm{Am}$ source. Secondly, a coincidence measurement was performed to determine the drift time of the charges from the point of their creation $t_\mathrm{init}$ to their arrival time $t_\mathrm{final}$ at the point contact. To this end, a collimated $^{22}\mathrm{Na}$ source, which emits two back-to-back gammas, was placed between the Ge detector and a NaI detector. The latter provides $t_\mathrm{init}$ and the PPC detector determines the arrival time $t_\mathrm{final}$. With this information the behavior of A/E as a function of the drift time was investigated. 

\begin{figure}
  \centering
  \subfigure[]{\includegraphics[width = 0.49\textwidth]{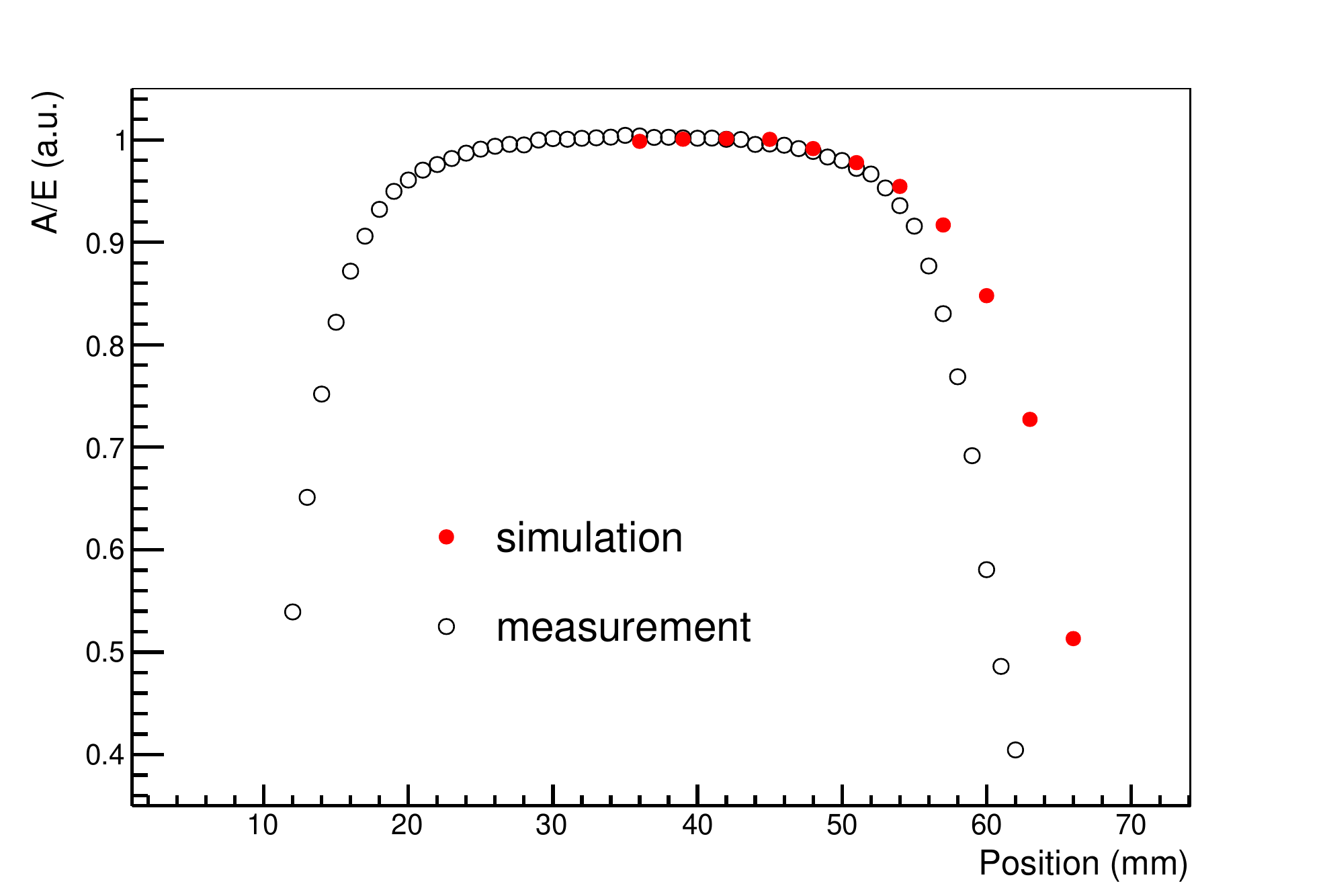}}
  \subfigure[]{\includegraphics[width = 0.49\textwidth]{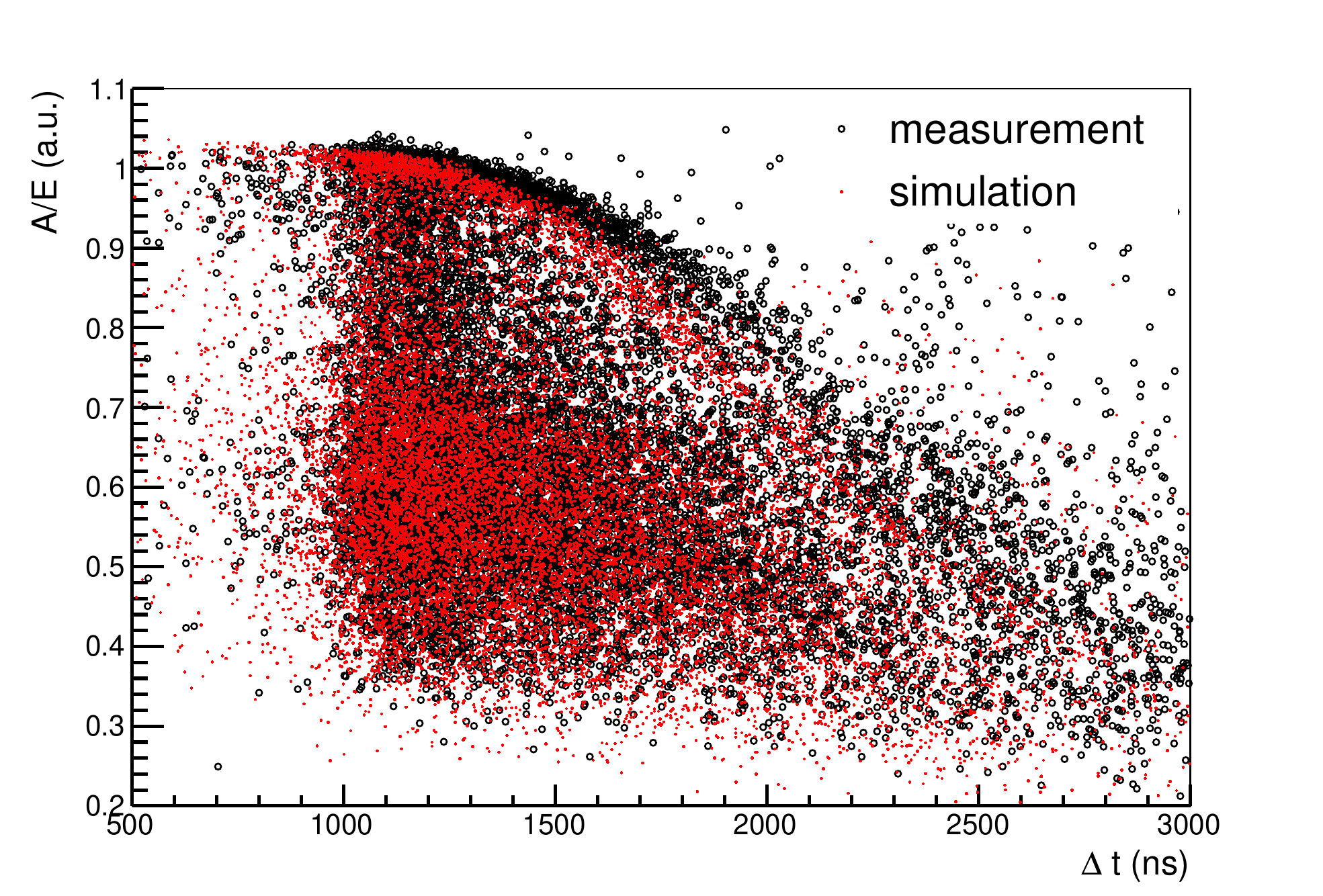}}
  \caption{(a): Comparison of measurement and simulation for an $^{241}\mathrm{Am}$ radial scan across the top of a ultra-low purity detector. Both measurement and simulation show a clear decrease of A/E at larger radii, corresponding to charges being created in the low-field corners of the detector. (b): Comparison of measurement and simulation of a coincidence measurement with a $^{22}\mathrm{Na}$ source. Again, the simulation reproduces the decreased A/E for large drift times $\Delta\mathrm{t}$, corresponding to charges created in the low-field regions. A constant cut in A/E would cut events at large drift times and large radii and thereby degrade the pulse-shape discrimination performance. For these simulations a detector with an impurity concentration of $1.6\cdot10^9$ net impurities per cubic centimeter at the top of the detector and an impurity gradient of $0.39\cdot10^9$ change in impurity density per centimeter was assumed.}
 \label{fig:lowimp}
\end{figure}

Both studies confirm that the maximum current A is reduced for charges created in the low-field corners of an ultra-low impurity detector, see Fig.~\ref{fig:lowimp}. This position dependence of the cut parameter is the reason for the degraded pulse-shape discrimination performance of ultra-low impurity detectors. Without the information on the position or the drift time (neither of which are provided by a non-segmented point-contact detector) the A/E cut removes events based on the creation position even if they are single-site in nature.

\section{Integration in M\small{AJORANA} mount}
\begin{figure}
  \centering
  \begin{minipage}{0.6\textwidth}
  \includegraphics[width = \textwidth]{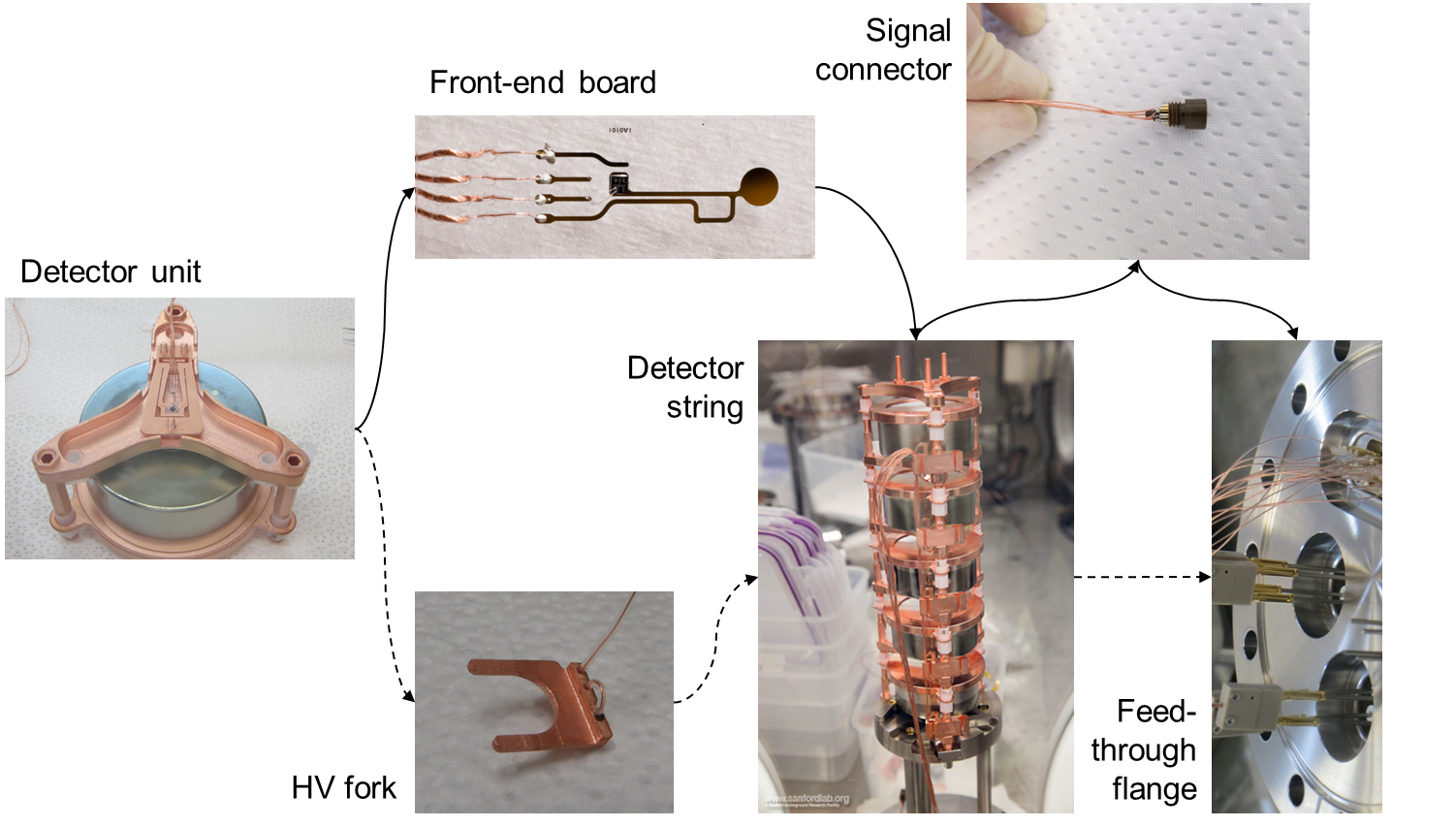}
  \end{minipage}
  \hfill
  \begin{minipage}{0.39\textwidth}
    \caption{Mounting of detectors in \textsc{Majorana} mount. The left photograph in the diagram shows a natural detector mounted in its electro-formed copper mount. The signal is extracted using a low mass front-end board (LMFE)~\cite{Taup}, HV is applied via the HV fork which is clamped between a PTFE nut and the HV ring at the bottom of the detector. 4--5 detector units are stacked to form a detector string. The HV line is directly brought to the feedthrough flange, whereas the signal line passes through an ultra-clean connector on top of the coldplate beneath which the string is attached.}
    \label{fig:assembly}
  \end{minipage}
\end{figure}

After completion of the characterization in the vendor cryostat, the detectors are removed from the vendor cryostat and mounted in \textsc{Majorana}'s low-background mount. The detector mounting structure is made from ultra-pure electro-formed copper and the front-end electronics~\cite{Taup}, which is placed in close vicinity to the detector, is made from carefully selected low-background material and with a minimal mass. Figure~\ref{fig:assembly} shows the integration of the detectors in the \textsc{Majorana} mounts. To closely pack the detectors in the final detector array, a combination of 4--5 natural and enriched detectors are stacked into a so-called detector string. Seven of these strings are placed in a module cryostat, two of which will be used to house and operate the \textsc{Majorana} detectors~\cite{MJD14}. The natural detectors used by \textsc{Majorana} are Broad Energy Ge (BEGe) detectors, produced by Canberra~\cite{canberra}. Their particular properties and performances are not detailed in this proceedings. 

Prior to the final installation of the strings in the module cryostats, they are again tested in the a so-called String Test Cryostat (STC). The goal of these tests is on the one hand to check the integrity of the detector, front-end electronics and HV connection after the re-installation of the detector in the \textsc{Majorana} mount, and on the other hand, this configuration allows the measurement of the crystal axis orientation of each detector in the string and the homogeneity of the dead layer along the side of each detector. The crystal axis orientation is of particular interest in view of a possible axion search with \textsc{Majorana}. 

\begin{figure}
  \centering
  \subfigure[]{\includegraphics[width = 0.49\textwidth]{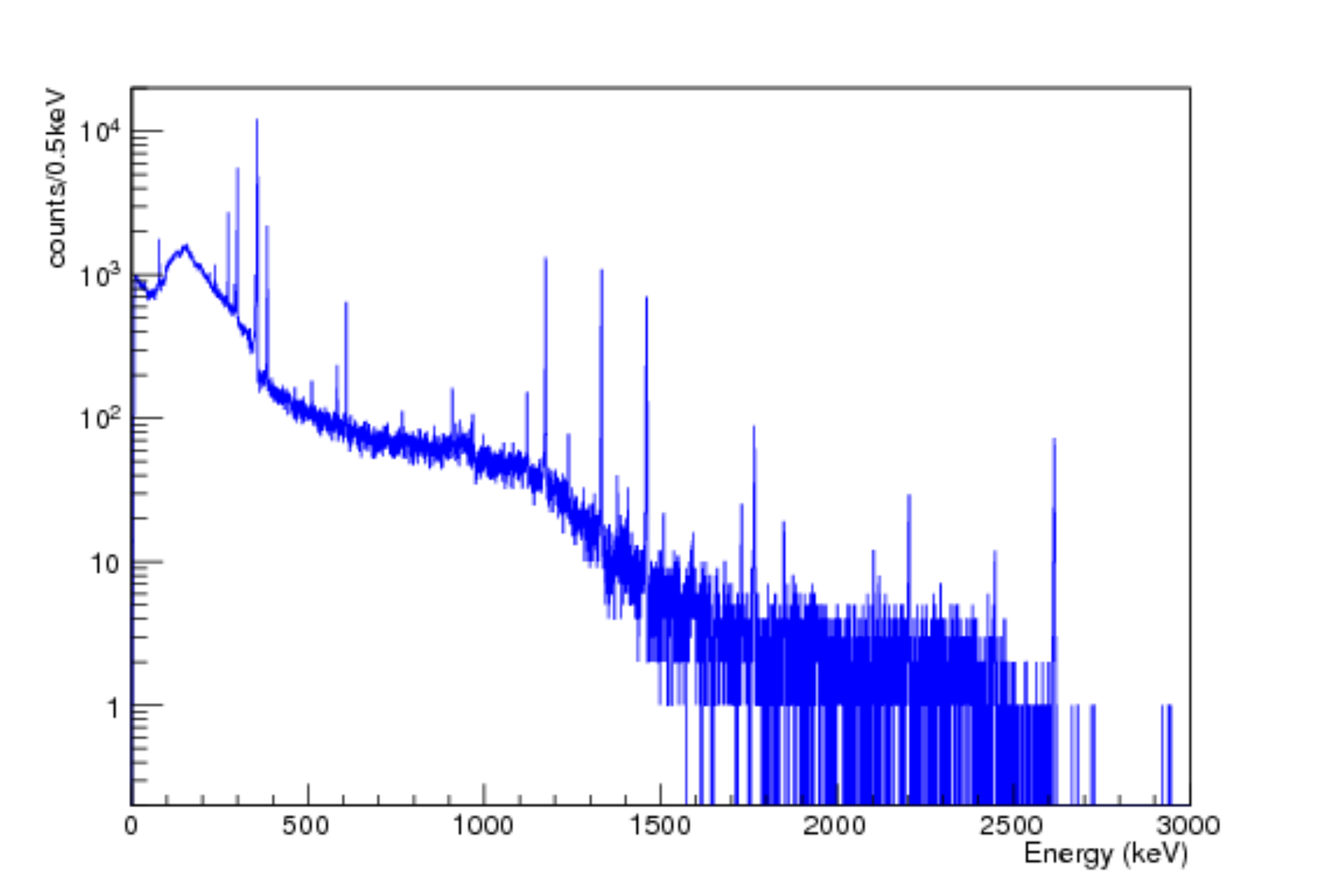}}
  \subfigure[]{\includegraphics[width = 0.49\textwidth]{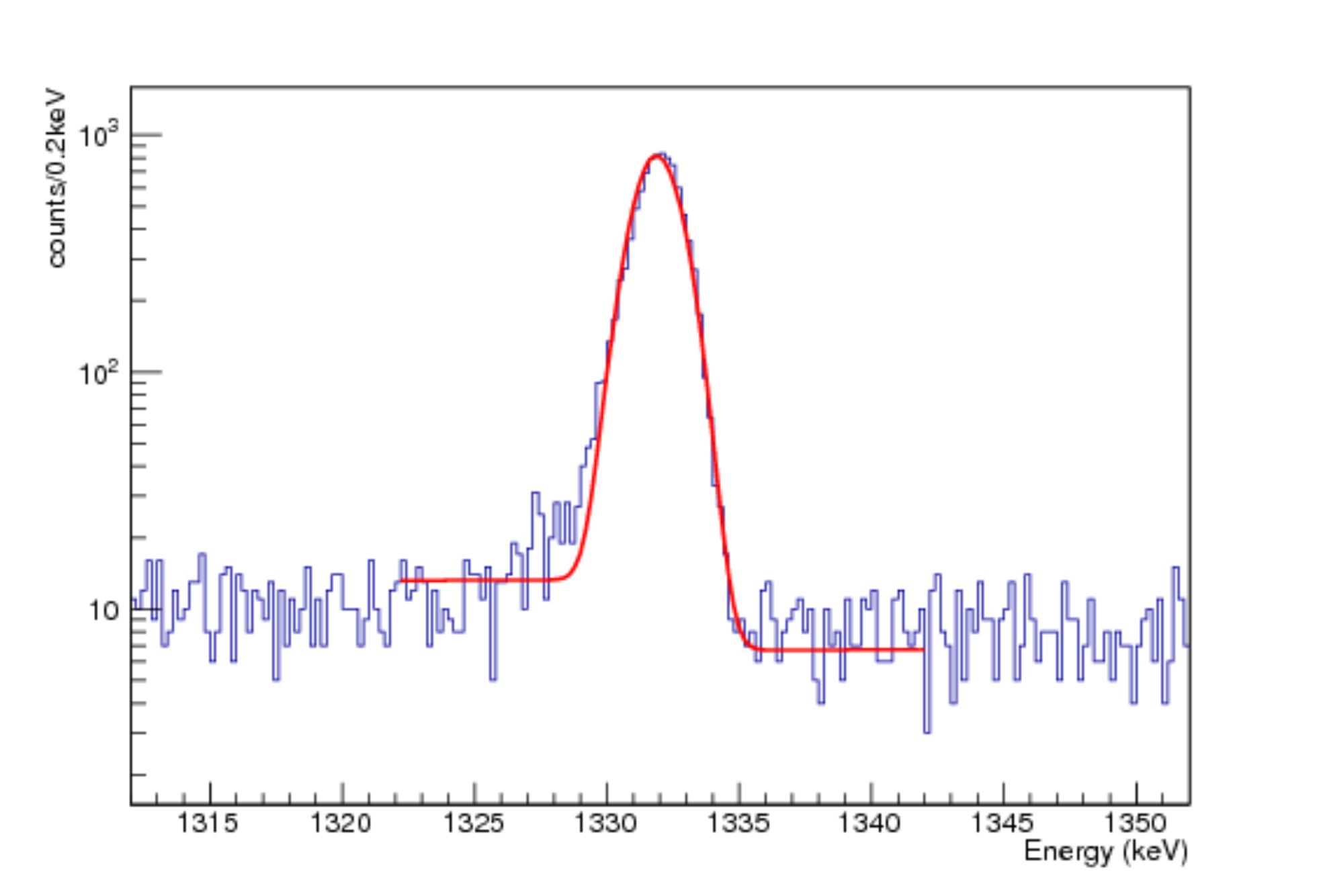}}
  \caption{(a): Calibration spectrum of a detector in a string test cryostat. (b): Zoom into the 1.33~MeV line of the ${}^{60}\mathrm{Co}$ calibration source. The energy resolution for this peak is FWHM = 1.9~keV. The small low energy tail is expected to be due to fact that this data was recorded without pole-zero correction.}
 \label{fig:STC}
\end{figure}

Fig.~\ref{fig:STC} shows a calibration measurement with a  ${}^{133}\mathrm{Ba}$ and  ${}^{60}\mathrm{Co}$ source taken with a detector operated in a STC. This measurement shows that the detector is operational at its standard operating voltage, i.e.\ the passivation layer has not been harmed in the installation process, the front-end electronics is intact, and the HV connection is functional. In this more complex setup (compared to the vendor cryostat) an energy resolution of $\mathrm{FWHM} \approx $2~keV at the 1.3~MeV  ${}^{60}\mathrm{Co}$-line could be achieved, which is similar to what was measured in the vendor cryostat and meets the \textsc{Majorana} requirements. 

\section{Conclusion}
\textsc{Majorana} experienced a good working relationship with ECP in Russia and ESI and ORTEC$^{\circledR}$ in the USA. The total sea-level exposure time for the enriched detectors could be kept better than the project specification of no more than 100 days. After reprocessing of scrap material \textsc{Majorana} expects ${\sim}$ 30~kg of enriched detectors.
All enriched detectors met requirements during characterization in the vendor cryostat at ORTEC$^{\circledR}$ and SURF. Dedicated simulations and measurements indicate that the slightly degraded pulse-shape discrimination behavior of one of the detectors is likely due to its ultra-low purity level.
At the time of this presentation \textsc{Majorana} is in the process of installing the detectors in specially designed low-background detector mounts and strings. First measurements in the string test cryostats show positive results with respect to integrity and performance of the detectors in the \textsc{Majorana} mounts.

\end{document}